\documentclass[12pt]{article} \usepackage{latexsym}
\usepackage{amsfonts,graphics} \usepackage{amsmath}

\normalbaselines 
\newcommand{\zRe}{E_R - i\frac{\Gamma_R}{2}}
\newcommand{\zR}{\ensuremath{E_R - i\Gamma_R/2}}
\newcommand{\HS}{\ensuremath{\mathcal{H}}}
\newcommand{\Phx}{\ensuremath{\Phi^\times}}
\newcommand{\la}{\rightarrow}
\newcommand{\bk}[2]{\ensuremath{\langle #1|#2 \rangle}}
\newcommand{\kt}[1]{\ensuremath{|#1\rangle}}
\newcommand{\br}[1]{\ensuremath{\langle #1|}}
\newcommand{\kb}[2]{\ensuremath{| #1\rangle\langle #2|}}
\newcommand{\pbk}[2]{\ensuremath{\langle {}^{+}#1|#2{}^{+} \rangle}}
\newcommand{\pkt}[1]{\ensuremath{|#1{}^{+}\rangle}}

\newcommand{\mbk}[2]{\ensuremath{\langle {}^{-}#1|#2^{-} \rangle}}
\newcommand{\mkt}[1]{\ensuremath{|#1{}^{-}\rangle}}

\newcommand{\pmkt}[1]{\ensuremath{|#1{}^{\pm}\rangle}}

\newcommand{\dbk}[2]{\ensuremath{( #1|#2 )}}
\newcommand{\dkt}[1]{\ensuremath{|#1 )}}

\begin{document}
\begin{center}
\vspace*{1.0cm}

{\LARGE{\bf Relativistic Resonances, Relativistic Gamow Vectors and
Representations of the Poincar\'{e} Semigroup\footnote{Plenary talk at
the International Symposium on Quantum Theory and Symmetries, July
1999, Goslar, Germany.  Written from transparencies and notes with
N.~L.~Harshman and M.~J.~Mithaiwala.}}} 

\vskip 1.5cm

{\large {\bf Arno R.~Bohm, N.~L.~Harshman and M.~J.~Mithaiwala}}

\vskip 0.5 cm 

Physics Department \\ 
The University of Texas at Austin\\  
Austin, Texas 78712, USA

\end{center}

\vspace{1 cm}

\begin{abstract}
The foundations of time asymmetric quantum theory are reviewed and are
applied to the construction of relativistic Gamow vectors.
Relativistic Gamow vectors are obtained from the resonance pole of the
$S$-matrix and furnish an irreducible representation of the
Poincar\'{e} semigroup.  They have all the properties needed to
represent relativistic quasistable particles and can be used to fix
the definition of mass and width of relativistic resonances like the
$Z$-boson.  Most remarkably, they have only a semigroup time evolution
into the forward light cone---expressing time asymmetry on the microphysical 
level.
\end{abstract}

\vspace{1 cm}

\section{Time Asymmetric Quantum Mechanics}

In classical physics one has time symmetric dynamical equations with
time asymmetric boundary conditions~\cite{PATO,Monster}.  These time
asymmetric boundary conditions come in
pairs: given one time asymmetric boundary condition, its
time reversed boundary condition 
can also be formulated mathematically.  For example in classical
electrodynamics one has retarded and advanced solutions of the time
symmetric dynamical (Maxwell) equations or in general relativity one
has time asymmetric big bang and big crunch solutions of
Einstein's time symmetric equation.  Except
for a few prominent cases of pedagogical importance (e.g. stationary
states or cyclic evolutions), the physics of our world is
predominantly time-asymmetric.  Somehow nature chooses one of the
pair of time asymmetric boundary conditions.

The standard quantum mechanics in Hilbert space~\cite{vonNeu} does not
allow time asymmetric boundary conditions for the 
Schr\"{o}dinger or von Neumann 
equation~\cite{Gleastone}.  However this is a consequence of the
\emph{mathematical} properties of the Hilbert space and need not imply that
quantum \emph{physics} is
strictly time symmetric.  It would be incredible if classical
electrodynamics had a radiation arrow of time and quantum
electrodynamics did not also have an arrow of time.

In quantum physics Peierls and Siegert considered many years
ago time asymmetric solutions with purely outgoing boundary
conditions~\cite{Peierls}.  The choice of appropriate dense
subspaces $\Phi_+$ and $\Phi_-$ of the (complete) Hilbert
space \HS\ allows the formulation of time asymmetric boundary conditions:
\begin{eqnarray}
\Phi_+ \subset \HS && \parbox[t]{3.25in}{for the out-states
$\{\psi^-\}$ of scattering theory which are actually observables
as defined by the registration apparatus (detector), and}\nonumber\\
\Phi_- \subset \HS && \parbox[t]{3.25in}{for the in-states
$\{\phi^+\}$ which are
prepared states 
as defined by the preparation apparatus (accelerator).}
\nonumber
\end{eqnarray}
Time asymmetric quantum theory distinguishes meticulously between
states $\{\phi^+\}$ and 
observables $\{\psi^-\}$.  Two different dense subspaces of the Hilbert
space \HS\ are chosen, $\Phi_-=\{\phi^+\}$ and $\Phi_+=\{\psi^-\}$.  The
standard Hilbert space quantum theory uses 
$\HS$ for both, $\{\psi^-\}=\{\phi^+\}=\HS$ and as a result is time
symmetric with a reversible unitary group time evolution.

In the theory of scattering and decay, a pair of time asymmetric
boundary conditions can be heuristically implemented by
choosing in- and out-plane wave ``states'' \pkt{E}\ and \mkt{E}\ which
are solutions of the Lippmann-Schwinger equation~\cite{Lippmann}
\begin{equation}
\pmkt{E} = \kt{E} + \frac{1}{E-H\pm i0}V\kt{E}=\Omega^\pm\kt{E},\label{LSeq}
\end{equation}
where $(H-V)\kt{E}=E\kt{E}$.
The energy distribution of the incident beam is given
by $|\pbk{E}{\phi}|^2 = |\bk{E}{\phi^{\mathrm{in}}}|^2$ and the energy
resolution of the detector (which for perfect efficiency is the energy
distribution of the detected ``out-states'') is measured as
$|\mbk{E}{\psi}|^2  =
|\bk{E}{\psi^{\mathrm{out}}}|^2$.

The sets $\{\pmkt{E}\}$ are the basis system that is used for the
Dirac basis vector expansion of the in-states $\phi^+\in\Phi_-$ and
the out-states (observables)
$\psi^-\in\Phi_+$:
\begin{eqnarray}
\psi^- &=& \sum_b \int^\infty_0 dE\, \mkt{E,b}\mbk{E,b}{\psi}\nonumber\\
\phi^+ &=& \sum_b \int^\infty_0 dE\, \pkt{E,b}\pbk{E,b}{\phi},
\end{eqnarray}
where $b$ are the degeneracy labels.
The Dirac kets of the Lippmann-Schwinger equation \pmkt{E}\ are in our time
asymmetric quantum theory~\cite{PRA99} antilinear functionals on the spaces
$\Phi_\mp$, i.e.~they are elements of the dual space
$\pmkt{E}\in\Phx_\mp$.

This leads to two rigged Hilbert spaces (RHS) and the
following new hypothesis for time asymmetric quantum theory\footnote{General states are
described by density operators $W^+=\sum_i w_i \kb{\phi_i^+}{\phi_i^+}$
and general observables by (positive definite) operators
$\Lambda^-=\sum_i \lambda_i \kb{\psi_i^-}{\psi_i^-}$.}:
\begin{eqnarray}
\parbox[c]{3in}{pure
registered observables or so-called ``out-states'' are described by
the vectors} && \psi^- \in \Phi_+\subset\HS\subset\Phx_+\nonumber\\ 
\parbox[c]{3in}{and pure prepared in-states are described by the
vectors}&&\phi^+ \in  \Phi_-\subset\HS\subset\Phx_-.\label{RHS's} 
\end{eqnarray}
This new hypothesis---with the appropriate choice for the spaces $\Phi_+$
and $\Phi_-$ given below in (\ref{caus})---is
essentially all by which our time asymmetric quantum theory differs
from the standard Hilbert space quantum mechanics which imposes
$\{\psi^-\}=\{\phi^+\}=\HS$ (or $\{\psi^-\}=\{\phi^+\}\subset\HS$).

In addition to the Dirac 
Lippmann-Schwinger kets $\pmkt{E}\in\Phx_\mp$, the dual spaces
$\Phx_\mp$ of
the RHS's also contain Gamow
kets $\pmkt{E_R \pm i\Gamma_R/2}\in\Phx_\mp$, which are generalized
eigenvectors of the (self-adjoint) Hamiltonian with complex eigenvalue
$(E_R \pm i\Gamma_R/2)$.  We use these Gamow kets to describe
quasistable particles.

We shall now mathematically define $\Phi_+$ and $\Phi_-$, and
therewith the RHS's
(\ref{RHS's}).  From a mathematical formulation of
causality expressed by the truism ``a state must be prepared before an
observable can be measured (registered) in it'', one can argue that
the energy wave functions \mbk{E}{\psi}\ are the boundary values of
analytic functions in the upper half energy plane (second sheet of the
$S$-matrix) and the \pbk{E}{\phi}\ are the same for the lower half
plane~\cite{PhysicaA97}.  Precisely, 
\begin{eqnarray}
\phi^+\in\Phi_- &\Leftrightarrow&
\pbk{E}{\phi}\in\mathcal{S}\cap\HS^2_-\nonumber\\
\psi^-\in\Phi_+ &\Leftrightarrow&
\mbk{E}{\psi}\in\mathcal{S}\cap\HS^2_+\label{caus},
\end{eqnarray}
where $\mathcal{S}$ is the Schwartz space and
$\mathcal{S}\cap\HS^2_\mp$ are well-behaved Hardy functions in the
lower (upper) half plane $\mathbb{C}^\mp$ of the second Riemann sheet
for the $S$-matrix $S(E)$.  The disparity between the labels $\pm$ for
the
vectors and the spaces (e.g.\ $\phi^+\in\Phi_-$) now makes sense.
The superscripts of the vectors is the standard notation of scattering
theory while the subscripts of the spaces comes from their
mathematical definition (\ref{caus}).  This correspondence
(\ref{caus}) between the physical state vectors and the mathematical
spaces is a wonderful example of
what Wigner calls ``the unreasonalbe effectiveness of mathematics in
the natural sciences''~\cite{Wigner2}.

\section{Gamow Vectors and Resonance States}

Stable states are described by bound state poles or by eigenvectors of
the self-adjoint Hamiltonian $H$ with real eigenvalue $E_n$:
\begin{eqnarray}
H^\dag\dkt{E_n} &=& E_n\dkt{E_n},\ \mbox{or equivalently}\nonumber\\
\dbk{Hf}{E_n} &=& E_n\dbk{f}{E_n}\ \mbox{for all}\ f\in\Phi\ \mbox{dense
in}\ \HS.\label{stable}
\end{eqnarray}
Quantum mechanical resonances are most commonly defined by the pair of
resonance poles in the second Riemann sheet of the analytically
continued $S$-matrix at the position $E_R \mp i\Gamma_R/2$; a decaying
state is associated to the pole at $z_R=\zR$.  Vector representations
of quasistable particles like the $K^0$ mesons in the Lee-Oehme-Yang
theory~\cite{LOY} use approximate methods such as the
Weisskopf-Wigner approximation~\cite{WWapp} for which ``there does not
exist...a rigorous theory to 
which these various methods can be considered approximations''~\cite{Levy}.

Often one hears the opinion that resonances and decaying
states are complicated objects and cannot be represented by simple
(exponentially decaying) state vectors.  This opinion differs from the
common practice to classify quasistable particles along with stable
particles~\cite{PDG} and also contradicts empirical facts: quasistable
particles are not qualitatively different from stable particles but
only quantitatively by a non-zero value of the width $\Gamma$ which
one takes to be the inverse lifetime, $\Gamma=\hbar/\tau$.  Stability
or the value of the lifetime is not a criterion for elementarity.  A
particle decays if it can and remains stable if selection rules for
some quantum numbers prevents it from decaying.

Stable and quasistable states should be described on the
same footing, e.g.~both defined by a pole of the $S$-matrix at the
position $E_n$ for stable particles and at the position $z_R=\zR$ for
quasistable particles. 
Then in analogy to the vector description (\ref{stable}) for stable
particles, there should also be a description of a quasistable particle
by a generalized eigenvector with eigenvalue $z_R$.  A vector
description is also needed to express the initial decay rate by the
Golden Rule as in \cite{LOY}.  Therefore, in analogy to (\ref{stable})
resonances and
quasistable particles will be described as generalized eigenvectors of
the self-adjoint Hamiltonian with complex eigenvalue $(\zR)$, where
$E_R$ represents 
the resonance energy and $\hbar/\Gamma_R$ is the lifetime: 
\begin{eqnarray}
\bk{H\psi^-}{\zRe {}^-}\equiv\br{\psi^-}H^\times\mkt{\zRe} &=&
(\zRe)\bk{\psi^-}{\zRe {}^-},\ \mbox{or}\nonumber\\ 
H^\times\mkt{z_R} &=& z_R\mkt{z_R},\label{unstable}
\end{eqnarray}
where the only distinction from (\ref{stable}) is that
$\bk{\psi^-}{z_R{}^-}$ are bra-kets ($\Phi_+$-continuous functionals)
and $\dbk{f}{E_n}$ are scalar products ($\HS$-continuous functionals).
We call the vectors $\mkt{z_R}\in\Phx_+$ Gamow vectors.

\begin{figure}
\resizebox{\textwidth}{!}{\includegraphics*[1.5in,5in][7in,7in]{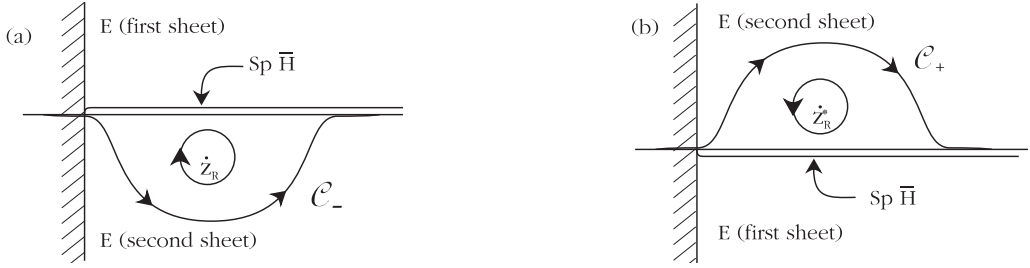}}
\caption{Deformation of the contour of integration from
$\mathrm{Sp\,\bar{H}}$ into the second sheet to include the resonance
pole.  The result of the contour deformation in Fig.~1b has been
mentioned in the text only in parenthesis.}
\end{figure}

To obtain this vector representation of quasistable states one
starts with the $S$-matrix element
\begin{eqnarray}
(\psi^{\mathrm{out}}, \phi^{\mathrm{out}}) & = & (\psi^{\mathrm{out}},
S\phi^{\mathrm{in}}) = (\phi^-, \phi^+) \nonumber \\
&=& \sum_{bb'}\int_0^\infty dE\, \bk{\psi^-}{b, E {}^-} \br{b} S(E) \kt{b}
\bk{{}^+b', E}{\phi^+}.\label{Smat}
\end{eqnarray}
and analytically continues the integral to the resonance pole
$z_R= E_R -i\Gamma/2$ of the $S$-matrix  $\br{b} S(z)
\kt{b}$~\cite{PhysicaA97} as shown in Fig.~1.  For 
this to be possible, the set of in-states $\{\phi^+\}\equiv\Phi_-$ and
out-states $\{\psi^-\}\equiv\Phi_+$ must have certain
analyticity properties.  In order to get a Breit-Wigner energy
distribution for the pole term of (\ref{Smat}), the energy wave functions
$\mbk{E}{\psi}$ and $\pbk{E}{\phi}$ must be well-behaved Hardy class
functions of the upper and lower half plane, respectively, in the
second sheet of the energy surface of the $S$-matrix.  This
is the same mathematical condition as provided by causality and stated
by (\ref{caus}).

Using the Cauchy formula, the analytically continued
kets $\mkt{E}\in\Phx_+$ become the Gamow kets $\psi^G =
\mkt{z_R}\in\Phx_+$ at the
resonance pole $z_R$.  For the Gamow kets one can prove the equations
(\ref{unstable})  The first equality of (\ref{unstable}) is the
definition of the conjugate operator 
$H^\times$ in $\Phx_+$, which is a unique extension of the
Hamiltonian $H^\dag=H$ in \HS.  The second equality of
(\ref{unstable}) says that
$\mkt{z_R}$ is a generalized eigenvector of $H$ with a complex
eigenvalue, similar to $H^\times\mkt{E} = E \mkt{E}$,
which in Dirac's notation is written $H\mkt{E}=E\mkt{E}$.

These Gamow kets $\mkt{z_R}$ have all the properties required of a
vector representing the ``state'' of an unstable particle or
(together with the ket $\pkt{z^*_R}=\pkt{E_R + i\Gamma_R/2}$ for the
$S$-matrix pole at $z_R^*$) of
a resonance in non-relativistic quantum physics.  Gamow vectors have
the following features:
\begin{enumerate}
\item They are derived as functionals of the resonance pole term at
$z_R=\zR$ (and at $z_R^*=E_R + i\Gamma_R/2$) in the second sheet of the
analytically continued $S$-matrix.

\item  The Gamow vectors have a Breit-Wigner energy distribution:
\begin{equation}
\bk{{}^-E}{\psi^G} = \frac{i\sqrt{\Gamma_R/2\pi}}{E - (E_R -
i\Gamma_R/2)},\ \ -\infty_{I\!I}<E<\infty.\label{BW}
\end{equation}

\item They have an asymmetric time evolution and obey an exponential law:
\begin{equation}
\psi^G(t)=e^{-iH^\times t}\mkt{z_R}=e^{-iE_R t}e^{-\Gamma_R t/2}
\mkt{z_R},\ \mbox{only for}\ t\geq 0.\label{semigroup}
\end{equation}

\item The decay probability of $\psi^G$ into the final non-interacting decay
products described by $\Lambda$, $\mathcal{P}(t) =
\mathrm{Tr}(\Lambda\kb{\psi^G}{\psi^G})$, can be calculated as a
function of time for $t\geq 0$, and from this the decay rate 
$R(t)=d\mathcal{P} / dt$ is obtained by differentiation.  This
leads to an exact Golden rule (with the natural lineshape given by a
Breit-Wigner) which in the Born approximation ($\psi^G\la f^D$;
$\Gamma_R/E_R\la 0$; $E_R\la E_0$,  where
$f_D$ is an
eigenvector of $H_0=H-V$ with eigenvalue $E_0$) goes into
Fermi's Golden rule No.~2 of Dirac.
\end{enumerate}

\section{Relativistic Resonances,\\ Relativistic Gamow Kets and\\
Poincar\'{e} Semigroup Representations}

Stable relativistic particles are defined and described by unitary
irreducible representations $\HS(s,j)$ of the Poincar\'{e} group
$\mathcal{P}$~\cite{Wigner} where $s$ is the eigenvalue of $P^\mu P_\mu$ and
$j$ is the spin.  In the irreducible representation space one commonly
uses the Wigner basis vectors \kt{\mathbf{p},j_3; (s,j)}, but one could
just as well use any other complete basis system, e.g. the 4-velocity
basis \kt{\hat{\mathbf{p}},j_3; (s,j)}\ where $\hat{p}_\mu =
p_\mu/\sqrt{s}$ and
$\kt{\hat{\mathbf{p}},j_3}=U(L(\hat{\mathbf{p}}))\kt{\hat{\mathbf{0}},j_3}$
with 
\begin{equation}
L^\mu_{\ \nu}(\mathbf{p})=L^\mu_{\ \nu}(\hat{\mathbf{p}}) =
\left( \begin{array}{cc} \frac{p^0}{m} & -\frac{p_j}{m} \\
\frac{p^i}{m} & \delta^i_{\ j} - \frac{\frac{p^m}{m}\frac{p_n}{m}}{1 +
\frac{p^0}{m}} \end{array} \right).
\end{equation}
This means every vector 
\begin{equation}
\phi^\pm\in\Phi_\mp \subset \HS(s,j) \subset
\Phx_\mp\label{vector}
\end{equation}
can be written as (Dirac basis vector expansion):
\begin{equation}
\phi^\pm = \sum_{j_3}\int \frac{d^3\hat{\mathbf{p}}}{2E}\kt{\hat{\mathbf{p}},
j_3 {}^\pm}\bk{{}^\pm\hat{\mathbf{p}}, j_3}{\phi^\pm}.
\end{equation}

Stable relativistic particles are characterized in addition to the
charge quantum numbers or particle labels by the value of mass $m$ and
spin $j$.  They are therefore described by the vectors $\phi^+(s=m^2,j)$ or
$\psi^-(s=m^2,j)$ of the unitary irreducible representation spaces
(\ref{vector}).  They lead to poles of the $j$th partial $S$-matrix
$S_j(s)$ at the
value $s=m^2$ on the real axis of the first sheet~\cite{Eden}.

Decaying states or 
resonances are defined by poles of the $j$th partial $S$-matrix
$S_j(s)$ on the second sheet at
\begin{subequations}
\begin{equation}\label{para}
s=s_R\equiv (m_R - i\frac{\Gamma_R}{2})^2 = m_\rho^2 - im_\rho
\Gamma_\rho,
\end{equation}
Where $(m_R,\Gamma_R)$ and $(m_\rho, \Gamma_\rho)$ are different
parameterizations of the complex value $s_R$.  Here
\begin{equation}
s\equiv (p_1 + p_2 + \cdots + p_n)^2 = E_R^2 - \mathbf{p}_R^2
\end{equation}
\end{subequations}
is the invariant mass squared, i.e.~the eigenvalue of the total mass
operator $P_\mu P^\mu$, $P^\mu=P^\mu_1 + P^\mu_2 + \cdots + P^\mu_n$.
The spin $j$ is the total angular momentum of
the $n$ decay products and is equal to the resonance spin $j=j_R$.

Quasistable relativistic particles are thus characterized by
$w_R\equiv\sqrt{s_R} = m_R - i\Gamma_R/2$ and by spin $j_R$, where we
call the resonance
mass $m_R$ and the resonance width $\Gamma_R$ for reason seen below in
(\ref{evol}).  More common is to call
$m_\rho$ and $\Gamma_\rho$ of (\ref{para}) the resonance mass and
width~\cite{PDG}.  
Some examples are:
\begin{enumerate}
\item Relativistic hadron resonances such as the $\rho$ meson in
$e^+e^-\la \rho \la \pi^+\pi^-$ with $(\Gamma/m)\sim 10^{-1}$.

\item Relativistic (weakly) decaying
states such as $K_s^0\la
\pi^+\pi^-$ with $(\Gamma/m)\sim 10^{-15}$.

\item The $Z$-bosons in $e^+e^-\la Z \la
\pi^+\pi^-$ with $(\Gamma/m)\sim 10^{-2}$.
\end{enumerate}

For these resonance scattering and decay processes one uses in the $S$-matrix
elements (\ref{Smat}), the relativistic Dirac Lippmann-Schwinger
``scattering states'' $\mkt{E, b}=\mkt{\hat{\mathbf{p}}, j_3; (s,j)}$
for ``physical'' values of $s$, or in other words for values of $s$
with $(m_1 + m_2 +
\cdots m_n)^2 \leq s < \infty$.
To obtain the relativistic Gamow kets one analytically continues these
relativistic Dirac Lippmann-Schwinger kets in the contour deformation
for the integral of (\ref{Smat}) in 
the following way: the real $s=p^2=(p_1 + p_2)^2$ becomes a complex
$s$ but the $\hat{p}_\mu=p_\mu/\sqrt{s}$ \emph{remain real}.
At the pole position $s=s_R$ one obtains the relativistic Gamow kets:
\begin{equation}
\mkt{\hat{\mathbf{p}},j_3; (s_R,j)} = \frac{i}{2\pi}
\int^{+\infty_{I\!I}}_{-\infty{I\!I}} ds\,\mkt{\hat{\mathbf{p}},j_3;
(s,j)} \frac{1}{s-s_R}.\label{rgk}
\end{equation}
These Gamow kets are basis vectors of a ``minimally complex''
semigroup representation of $\mathcal{P}$ which is characterized by
$(s_R, j)$.  This is the analogy of the representation for
non-relativistic Gamow vectors~\cite{PhysicaA97}:
\begin{equation}
\mkt{z_R} = \frac{i}{2\pi}\int_{-\infty_{I\!I}}^{+\infty_{I\!I}} dE
\mkt{E} \frac{1}{E - z_R}.
\end{equation}
The relativistic Gamow vectors have according to (\ref{rgk}) a
``relativistic Breit-Wigner'' $s$-distribution given by $(s-(M_R -
i\Gamma/2)^2)^{-1}$.  They can be 
shown to be generalized eigenvectors of the invariant mass-squared
operator $M^2=P_\mu P^\mu$~\cite{LANL9911}, i.e.\ they fulfill
\begin{eqnarray}
\langle \psi^-|(M^2)^{\times}|j,s_R;b^- \rangle &=& \frac{i}{2\pi}
\int_{-\infty}^{+\infty} ds\, s \langle \psi^-|j,s;b^- \rangle
\frac{1}{s-s_R}\nonumber\\
& =& s_R \langle \psi^-|j,s;b^- \rangle\,
\label{eq:action.of.M2} 
\end{eqnarray}
for every $\psi^- \in
\Phi_+ \subset \HS(s_R, j) \subset \Phi_+^\times$.

The Lorentz transformations $\Lambda$ in these minimally complex
representations are represented by a group of unitary operators
$U(\Lambda)$ and act in the well-known way:
\begin{equation}
U(\Lambda)\mkt{\hat{\mathbf{p}},j_3; (s_R,j)}=
\sum_{j_3'}D^j_{j_3'j_3}(\mathcal{R}(\Lambda, \hat{\mathbf{p}}))
\mkt{\Lambda \hat{\mathbf{p}},j_3'; (s_R,j)}.
\end{equation}
However the space-time translations can no more be represented by a
group of operators $U(a, 1)$ in
\begin{equation}\label{phiplus}
\Phi_+ \subset \HS(s_R, j) \subset \Phi_+.
\end{equation}
The rigged Hilbert spaces of Hardy class have to be employed in
rigorously obtaining (\ref{rgk}) and (\ref{eq:action.of.M2}) and then
one can see that only the semigroup of space-time 
translations into the forward light cone can be represented in the
space (\ref{phiplus}).  Much
like in the RHS theory of non-relativistic Gamow vectors, one can show
that the time translation of the decaying state in the rest frame
$\hat{\mathbf{p}}=\mathbf{0}$ is given by
\begin{equation}\label{evol}
e^{-iHt}\mkt{\mathbf{0},j_3; (s_R,j)}=e^{-im_Rt}
e^{-\Gamma_Rt/2} \mkt{\mathbf{0},j_3; (s_R,j)},\ t\geq 0\ \mbox{only},
\end{equation}
where $t$ is the time in the rest system.

Thus, the relativistic Gamow vectors have a semigroup time evolution
and obey the exponential law.  From (\ref{evol}) it follows that the
lifetime of the particle represented by the relativistic Gamow vector
is given by  
$\tau_R=\hbar/\Gamma_R$ where the width $\Gamma_R$ is the
$\mathrm{Im}\sqrt{s_R}$ given by the resonance pole position $s_R$ of
the relativistic $S$-matrix.  The real resonance mass $m_R$ in
(\ref{evol}) is the $\mathrm{Re}\sqrt{s_R}$ and therefore not exactly
the same as the peak position $m_\rho\equiv
m_R\sqrt{1-1/4(\Gamma_R/m_R)^2}$ of the modulus of the relativistic
Breit-Wigner amplitude of (\ref{rgk}).

\section{Summary}

Our fundamental assertion is that stable and unstable particles are
not fundamentally different.  The theoretical treatment of stable and
unstable particles should therefore be the same.  Within the Hilbert
space (HS) quantum theory, time asymmetric 
boundary conditions cannot be formulated and a theory of decaying
states within the HS framework can only be approximate~\cite{Levy}.
Within the RHS, time 
asymmetric quantum theory can be introduced by the new hypothesis
(\ref{RHS's},\ref{caus}) which
allows a distinction between states and observables.

Within this theory, Gamow vectors, both relativistic and
non-relativistic, satisfy all necessary properties of representing
quasi-stable states and resonances.  They are associated to poles of
the $S$-matrix, have Breit-Wigner energy distribution and obey
an exponential semigroup time evolution.  All of these properties can
be formulated mathematically precisely, and a particular property of
the Gamow vector is that it relates the width $\Gamma_R$ of the
Breit-Wigner energy
distribution and the lifetime $\tau$ of the exponential
decay $\tau=\hbar/\Gamma_R$.  Without the Gamow vectors, the width and
inverse lifetime only can be shown to be approximately equal by use of
some approximate method based on Weisskopf-Wigner~\cite{WWapp}.
Consequently, one did not know whether to choose $(m_R, \Gamma_R)$ or
$(m_\rho, \Gamma_\rho)$ of (\ref{para}) or any other $(m,\Gamma)$ as
the mass and width of relativistic resonances.

The relativistic Gamow vectors
unify in a fundamental picture both stable and unstable relativistic particles.
Both are given by irreducible representations of Poincar\'{e} transformations.
Stable particles are unitary representations characterized by a real mass and
have unitary group
time evolution,  
quasi-stable particles are ``minimally complex''
semigroup representations characterized by 
a complex mass and have
semigroup time evolution with an arrow of time.

That relativistic Gamow states $\mkt{z_R}$ possess only semigroup
transformations 
into the forward light cone (and that others possess only those into
the backward light cone) and no space-like translations (for either
case) is a result 
whose interpretation is not yet clear.  It is the relativistic
analogue of the semigroup time evolution in non-relativistic quantum
mechanics which can be understood as the causality condition that a
state must be first prepared before an observable can be measured in
it~\cite{PRA99}.  This time asymmetry on the microphysical
level---irreversibility without entropy increase and without violation
of 
time reversal invariance---was the most surprising and remarkable
property of the non-relativistic and relativistic Gamow vectors.

\section*{Acknowledgment} 
These lecture notes are dedicated to Professor H.~D.~Doebner on the
occasion of his graduation to emeritus status.  We gratefully
acknowledge valuable support from the Welch Foundation.

\end{document}